\documentclass{ab}

\usepackage{graphicx}
\usepackage{natbib}
\usepackage{lscape}

\newcommand{\kms}{$\textrm{km~s$^{-1}$}$}

\newcommand{\pak}{PA$_{kin}$}

\def\degr{\hbox{$^\circ$}}
\def\arcmin{\hbox{$^\prime$}}
\def\arcsec{\hbox{$^{\prime\prime}$}}

\begin{document} 

\title{Gas Kinematics in the Magellanic-Type Galaxy NGC 7292}
       
\author{A.~S.~Gusev,$^1$ 
        A.~V.~Moiseev$^{1,2}$
        and S.~G.~Zheltoukhov$^1$}

\institute{$^1$ Sternberg Astronomical Institute, Moscow State University, Moscow, 119234 Russia \\
           $^2$ Special Astrophysical Observatory, Russian Academy of Sciences, Nizhnii Arkhyz, 
           369167 Russia}

\date{Received April 22, 2023; revised June 5, 2023; accepted June 6, 2023}
\offprints{Alexander~S.~Gusev, \email{gusev@sai.msu.ru}; 
           Alexei~V.~Moiseev, \email{moisav@gmail.com}}

\titlerunning{Gas Kinematics in the Magellanic-Type Galaxy NGC 7292}
\authorrunning{Gusev et al.}

\abstract{The paper presents results of studying the kinematics of the ionized gas in the galaxy 
of the Large Magellanic Cloud type NGC 7292 obtained with the 2.5-m telescope of the Caucasian 
Mountain Observatory (CMO SAI MSU) and the 6-m BTA telescope of the Special Astrophysical Observatory 
(SAO RAS). Analysis of the velocity ﬁelds of the ionized and neutral hydrogen showed that the 
kinematic center of NGC~7292 located at the center of the bar, northwest of the photometric center 
of the galaxy (the southeastern end of the bar) previously taken as the center of NGC~7292. In 
addition to the circular rotation of the gas, the radial motions associated with the bar play 
a significant role in the kinematics of the disk. The observed perturbations of the gaseous-disk 
kinematics induced by the ongoing star formation do not exceed those caused by the bar. It is 
possible that part of the non-circular motions (at the southeastern end of the bar which is the 
brightest HII~region) may be related to the effects of the capture of a dwarf companion or a gaseous 
cloud. \\

                
{\bf DOI:} 10.1134/S1990341323700104 \\
}

\maketitle

\section{INTRODUCTION}
\label{sect:intro}

\begin{figure*}
%\vspace{9mm}
\centerline{\includegraphics[width=17.8cm]{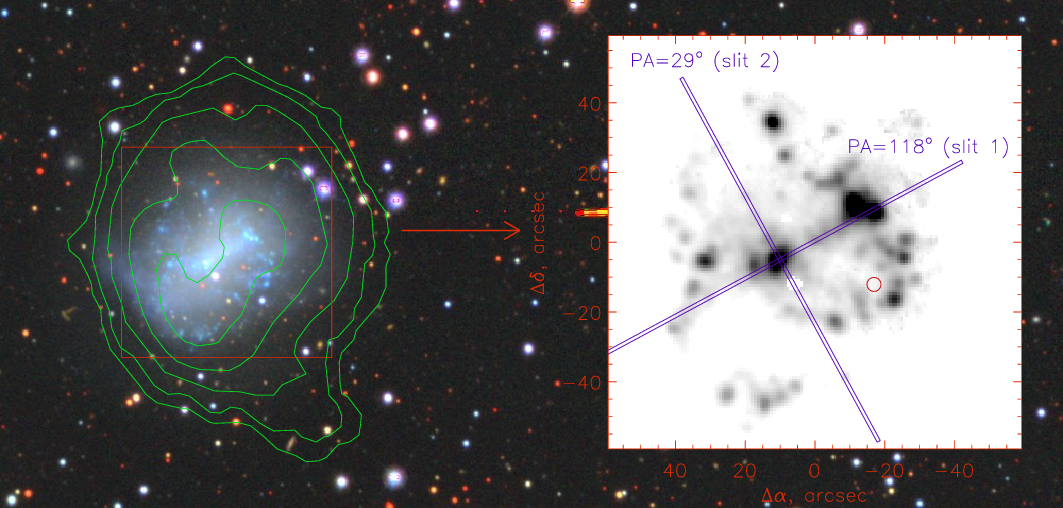}}
\caption{Image of NGC~7292 from the DESI Legacy survey \citep{legacysurvey}. The HI lines of equal 
density are shown in green according to \cite{biswas2022}. In the right-hand inset, the image 
in the H$\alpha$ line obtained with the scanning FPI is given. We show the slit positions of the 
spectrograph. The center of coordinates corresponds to the nucleus of the galaxy. The red circle 
indicates the position of the supernova 1964H. North is on the top, East is on the left.
\label{fig:map1}}
\end{figure*}

Disk galaxies of extremely late types of intermediate masses, the closest analogue of which is 
the Large Magellanic Cloud, are a relatively rare type of galaxies. Being, in most cases, of an 
asymmetric shape, they are the result of complex evolutionary processes: interactions with large 
neighbours or possible merging with companions 
\citep[see, e.g.][]{harris2009,besla2012,yozin2014,koch2015,pardy2016,siejkowski2018}. One of 
these star systems is a nearby but relatively poorly-studied galaxy of the IBm type 
(according to the NED\footnote{http://ned.ipac.caltech.edu} database) NGC~7292 (Fig.~\ref{fig:map1}).

The galaxy is prominent with a heavy bar displaced from the disk. At the end of the bar, large 
star-forming regions are located, which are  the brightest areas in the short-wavelength optical 
bands ($U$ and $B$) and in the H$\alpha$ line (Fig.~\ref{fig:map1}). The southeastern part of the bar 
is usually taken as the galaxy's center (see, for example, the NED database), although, 
\cite{gusev1996}, based on multicolour surface photometry data, assumed that the center of the bar 
is the center of NGC~7292.

The parameters of the galaxy are determined quite reliably. According to the LEDA 
database\footnote{http://leda.univ-lyon1.fr}, its absolute magnitude $M(B)^i_0=-16.7\pm1.0$, 
the positional angle is $113\degr$, and the inclination of the disk equals $54.4\degr$. The bar, 
which suppresses the emission of a weaker disk, plays a decisive role in the calculation of the 
last two parameters. The distance to NGC~7292 also remains an open question (see the distance 
estimates in NED), while the radial velocity of the galaxy relative to the Sun (986~\kms) is 
determined quite reliably. We accepted the distance to the galaxy as $d=6.82$~Mpc according to 
\citet{tully2009} which corresponds to the 33~pc\,arcsec$^{-1}$ scale. All the distance-dependent 
estimates taken from other papers (luminosity, sizes, etc.) are given precisely to this distance 
in our paper.

The galaxy NGC~7292 belongs to the NGC~7331 group but is located on its periphery and has no 
close companions \citep{ludwig2012}.

The rotational curve of NGC~7292 was obtained in the paper by \cite{esipov1991} from the [SII] 
emission lines. With an accepted disk inclination of $36\degr$, the authors have found that the 
rotation velocity reaches 100~\kms at a distance of 800~pc from the center. However, in the specified 
paper, there is no information on the kinematic center and the positional angle of NGC~7292 which 
does not allow one to verify the results obtained by \cite{esipov1991}.

Recent studies of the neutral hydrogen carried out with the Indian GMRT radio interferometer 
\citep{biswas2022} have shown that NGC~7292 has a regularly rotating disk HI of an approximate 
diameter of $\approx3\arcmin$ (6~kpc) and a mass of $(2.08\pm0.06)\times10^8M_{\odot}$ (at the HI 
column density level from $1\times10^{20}$~cm$^{-2}$ and higher). The authors give the 
''photometric'' positional angle of the disk equal to $9\degr$, however, the HI velocity field map 
indicates the kinematic positional angle \pak$\approx250\degr$.

The HI disk in NGC~7292 is round with the exception of a small protrusion on the southwestern 
outskirts (see Fig.~\ref{fig:map1}). This protrusion does not differ from the major HI disk in 
velocity and velocity dispersion \citep[][see details in Section~\ref{sect:res}]{biswas2022}. 
The observed asymmetric structure is morphologically similar to the part of the tidal tail 
which, together with the distorted shape of the outer optical isophotes, can be indicative of 
a relatively recent (less than 1-2 disk revolutions ago) event associated with the capture of 
the outer matter.

A large number of HII regions are observed in the galaxy \citep{gusev2021}. A supernova of type~II 
(1964H) is known that burst in the western chain of HII regions \citep[the J2000 coordinates: 
$\alpha = 22^h28^m24.06^s$, $\delta = +30\degr17\arcmin23.3\arcsec$;][]{crowther2013}.

Spectroscopic observations with the bar-aligned long slit (PA $= 118.3\degr$) carried out in the 
paper by \cite{gusev2021} allowed one to estimate the chemical abundance of the gas in nine HII 
regions. It has been shown that the oxygen and nitrogen abundances in the HII regions are 
typical of galaxies of the speciﬁed luminosity (O/H $=8.26\pm0.03$~dex, N/H $= 6.86\pm0.07$~dex) 
with the absence of a radial metallicity gradient. A weak gradient of N/O was noticed along the 
bar major axis of NGC~7292: the N/O ratio is decreasing from the eastern to the western part of 
the galaxy.

\begin{figure*}
%\vspace{9mm}
\centerline{\includegraphics[height=9cm]{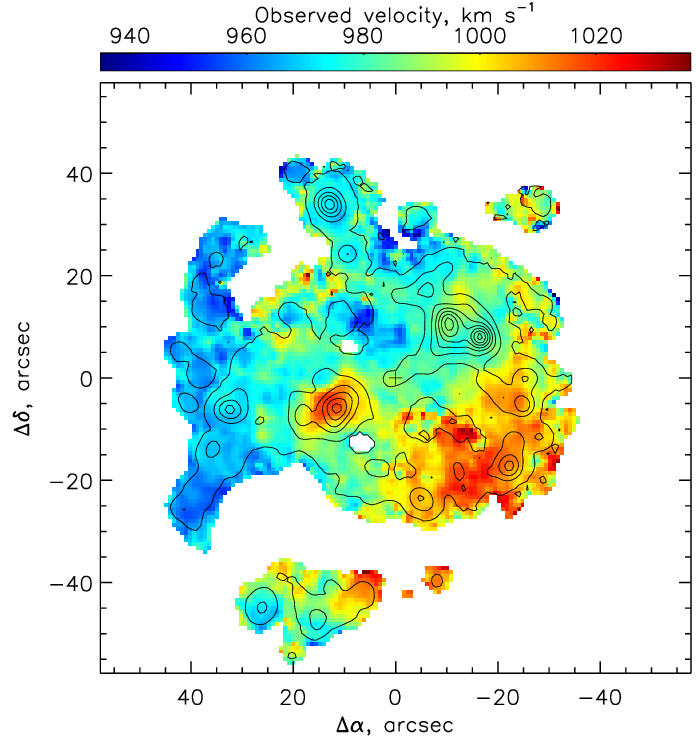}
\includegraphics[height=9cm]{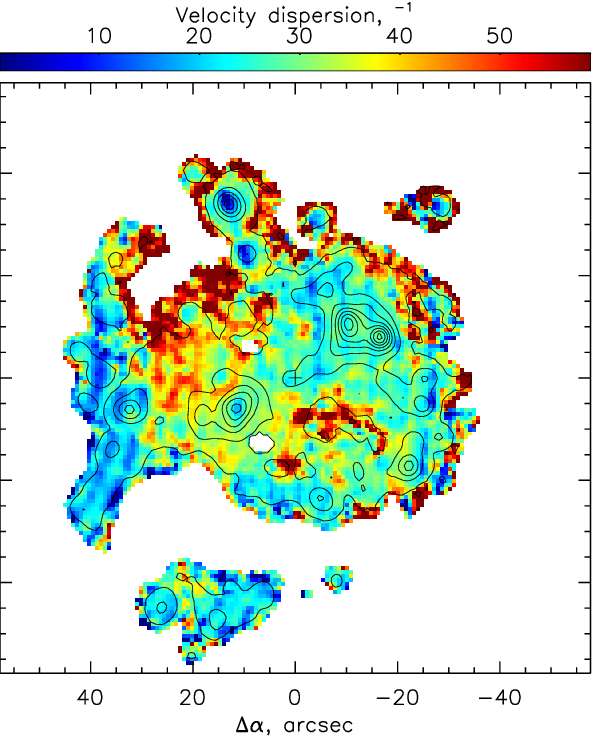}}
\centerline{\includegraphics[height=9cm]{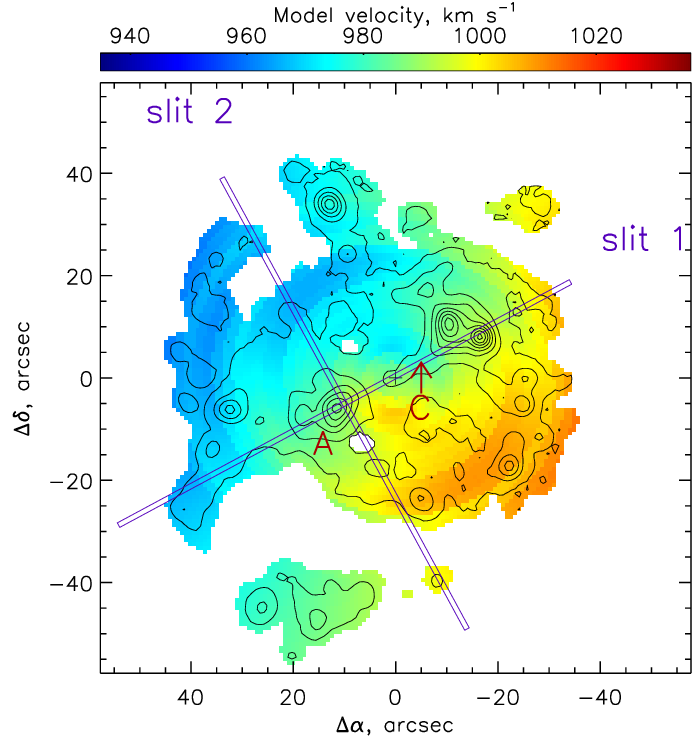}
\includegraphics[height=9cm]{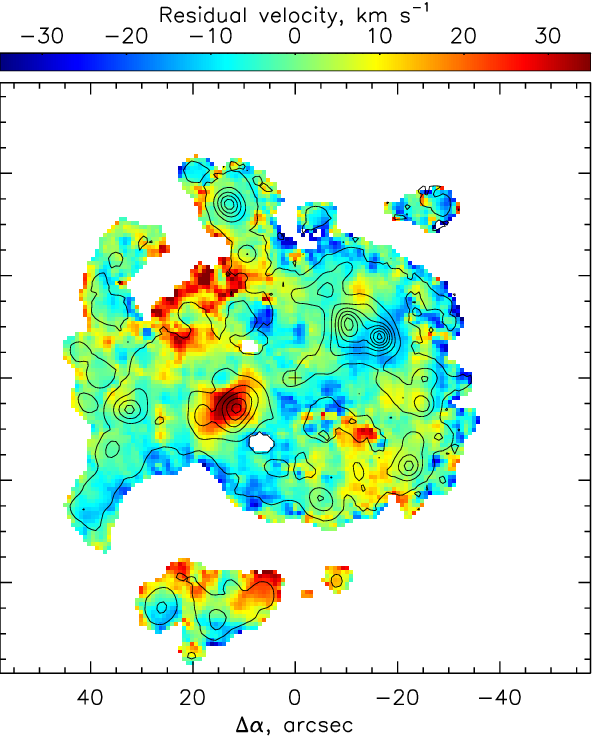}}
\caption{Ionized-gas kinematics in the H$\alpha$ line from the 6-m telescope data. The upper row 
shows the observed parameters: the radial velocity field (top-left) and the velocity dispersion field 
(top-right). The lower row shows the model velocity field (bottom-left) and the distribution of the 
deviation from the model (bottom-right). The cross in the maps marks the kinematic center of the 
galaxy. The arrow points the position of region C in the bar center. In the panel of the model 
velocity field, the positions of the spectrograph slits are indicated and two HII regions discussed in 
the text are marked.
\label{fig:ifpmaps}}
\end{figure*}

Among the HII regions, a region in the bar center stands out (region C in Fig.~\ref{fig:ifpmaps}), 
in which a low N/O ratio is observed; in the diagnostic diagram [NII]/H$\alpha$ -- [OIII]/H$\beta$ 
\citep{baldwin1981,kewley2001,kauffmann2003}, it is located near the boundary separating objects 
with thermal and non-thermal (shock waves, ionizing emission from the active galactic nucleus) 
mechanism of excitation of emission lines \citep[see fig.~5 in the paper by][]{gusev2021}. 
In addition, the region at the center of the bar is characterized by the maximum O/H ratio in 
the galaxy. The brightest HII region located at the southeastern end of the bar (region A in 
Fig.~\ref{fig:ifpmaps}) shows the maximum N/O value with the average O/H \citep{gusev2021}.

The spectrum of the galaxy obtained in the paper by \cite{gusev2021} attracted attention due to 
the unusual rotation curve which served an incentive to studying the gas kinematics in NGC~7292 
in detail.

\section{OBSERVATIONS AND DATA REDUCTION}

\subsection{Panoramic Spectroscopy with the Fabry-Perot Interferometer}
\label{sect:obs2}

Observations of NGC~7292 with the scanning FPI installed in the SCORPIO multi-mode focal reducer 
\citep{AfanasievMoiseev2011} were carried out with the 6-m SAO RAS telescope on October 9/10, 
2021. The spectral range around the H$\alpha$ line was separated with the $FWHM\approx14$\AA\, 
narrow-band filter. During the observations, the gap between the plates of FPI was varied so that 
the obtained 40 interferograms uniformly filled the entire overlap-free spectral range 
($\Delta\lambda=36$\AA). The exposure time was 90~s with the quality of stellar images being 
$1.8-2.8\arcsec$. The spectral resolution of FPI was 1.7\AA\, (which corresponded to 
78~km\,s$^{-1}$ at the H$\alpha$ wavelength). The detector, the E2V 261-84 CCD camera, provided 
the field of view of about $6.6\arcmin$ with a discretization of $0.78\arcsec$/px in the readout 
mode with the $2\times2$ hardware binning.

\begin{table}
\begin{center}
\caption{Observation log of CMO of SAI MSU.}
\label{tab:obs1}
\begin{tabular}{|l|c|r|r|r|}
\hline\hline
Slit & Date & Exp, s & PA, degr & $\delta\lambda_{\rm lab}$, \AA \\
\hline
1  & 10/11.12.2020 & 100+         &   118.3 & 2.53$\pm$0.12 \\
   &               & 3$\times$900 &          &  \\
2a & 01/02.05.2021 & 900          & --151.5 & 2.61$\pm$0.19 \\
2b & 02/03.05.2021 & 3$\times$900 &    28.5 & 2.53$\pm$0.14 \\[1mm]
\hline
\end{tabular}
\end{center}
\end{table}

After the primary reduction \citep[a detailed description of the software used and references to 
original papers are given in][]{Moiseev2021AstBu..76..316M}, the observations were presented as 
a data cube, in which each pixel in the field of view contained the spectrum of the region around 
the H$\alpha$ line. The astrometric calibration was carried out online with astrometry.net 
\citep{astrometry.net}. The resulting angular resolution after smoothing during the primary 
reduction was $2.8\arcsec$.

The brightness distributions in the H$\alpha$ emission, the radial velocity field in this line, 
and the field of the radial velocity dispersion free from instrumental broadening 
(Fig.~\ref{fig:ifpmaps}) were constructed by approximating the spectra with the Voigt profile 
according to the method described in the paper by \cite{Moiseev2008AstBu..63..181M}.

\subsection{Long-Slit Spectroscopic Observations}

Observations were carried out with the 2.5-m telescope of the Caucasian Mountain Observatory of 
SAI MSU using the TDS transient twobeam spectrograph (TDS -- the transient two-beam spectrograph). 
The spectrograph simultaneously operates in two channels: the blue (in the range of 3600--5770\AA\, 
with a dispersion of 1.21\AA/px, and a spectral resolution of 3.6\AA) and red (in the range of 
5673–7460\AA\, with a dispersion of 0.87\AA/px, and a spectral resolution of 2.6\AA). Two CCDs use 
the E2V 42-10 detectors of the size of $2048\times512$~px$^2$. The pixel size is $0.363\arcsec$, 
the size of the used slits is $180\arcsec\times 1\arcsec$. You can see the detailed description 
of the spectrograph the paper by \cite{potanin2020}.

The observations were carried out in December 2020 and May 2021 (see the observation log in 
Table~\ref{tab:obs1}) with the image quality of $1.2-1.8\arcsec$. During the first observation 
set, the slit was positioned along the bar of the galaxy (PA $= 118.3\degr$), and during the 
second -- perpendicular to the first (PA $= 28.5\degr$), so that the slit passed through the 
brightest region of NGC~7292, the eastern end of the bar (see Fig.~\ref{fig:map1}). During the 
observations in May, the slit in the second night (set~2b) was turned by $180\degr$ with respect 
to the first night (set~2a). The instrumental spectral line width (FWHM) $\delta\lambda_{\rm lab}$, 
determined from the emission lines of the night sky, was $2.5-2.6$\AA\, (115-120~km\,s$^{-1}$, 
see Table~\ref{tab:obs1}).

The observation procedure included obtaining the flat fields and calibration images in the 
beginning and at the end of each observation series. Spectrophotometric standards were observed 
directly after the observations of the galaxy with the same air mass.

The original data was reduced according to the standard procedure including the dark-current 
correction, cosmic ray removal, flat-field correction, wavelength calibration using the Ne-Al-Si 
standard lamp and its correction with the night-sky lines, photometric calibration, adding the 
spectra of individual exposures, subtracting the background, and converting to the one-dimensional 
spectra \citep[see the paper by][for detail]{potanin2020}. For the primary data reduction, we used 
the data reduction package based on Python developed in SAI MSU by A.~V.~Dodin; the further 
reduction was fulfilled with the ESO-MIDAS system for data reduction.

The radial velocity and velocity dispersion were determined from the wavelength and the H$\alpha$ 
line width in $1\arcsec$ increments. For this purpose, the spectrum was integrated within 
$1\arcsec$, while the wavelength and the H$\alpha$ line half-width (in the presence of emission) 
were determined by approximating the line with a Gaussian. The spectra were not corrected for the 
absorption component from the stellar population. As has been shown in the paper by 
\cite{gusev2021}, its contribution to the H$\alpha$ lines is negligible: in all the HII regions 
studied, EW(H$\alpha$) $>40$~\AA.

The observed radial velocity is reduced to heliocentric. The velocity dispersion $\sigma$ was 
calculated taking into account the instrumental width $\delta\lambda_{\rm lab}$: 
$\sigma^2 = \sigma_{\rm obs}^2 - \sigma_{\rm lab}^2$, where 
$\sigma_{\rm lab} = \delta\lambda_{\rm lab}/(2\sqrt{2\ln2})$, 
$\sigma_{\rm obs} = \delta\lambda_{\rm obs}/(2\sqrt{2\ln2})$.

\section{RESULTS}
\label{sect:res}

\subsection{Radial Velocities and Velocity Dispersion of HII}

\begin{figure}
\vspace{3mm}
\centerline{\includegraphics[width=8.5cm]{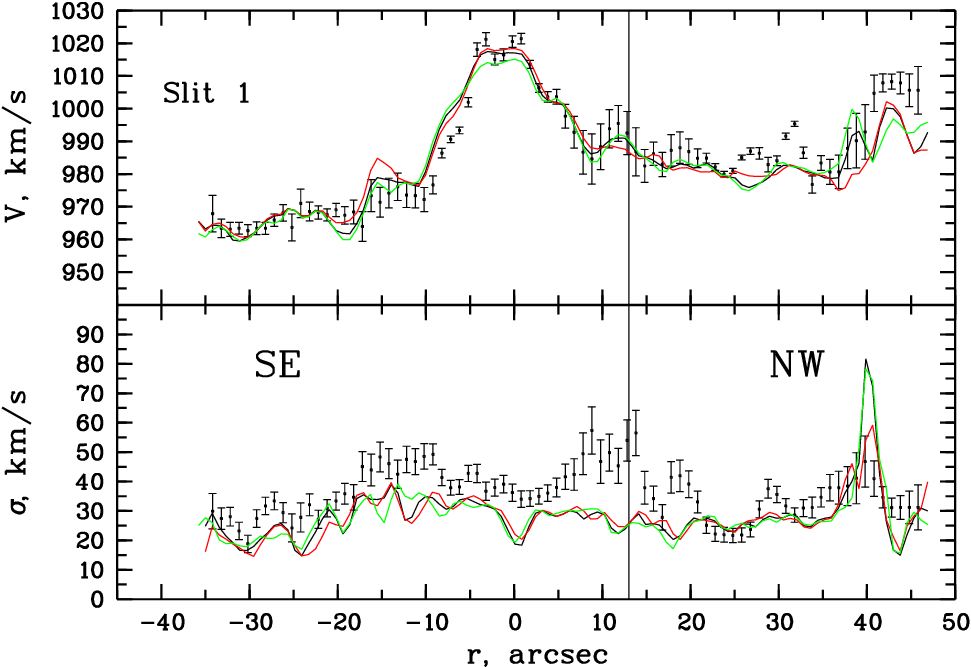}}
\caption{Radial velocity (top) and velocity dispersion (bottom) of HII along the first slit position 
(PA $= 118.28\degr$) according to the long-slit spectroscopy (the dots) and panoramic spectroscopy 
(the black curves). The red and green curves show the $V$ and $\sigma$ profiles obtained with the 
Fabry-Perot interferometer and right–left shifted relative to the slit by $1\arcsec$. Measurement 
errors are shown. The center of the brightest HII region (the southeastern end of the bar, region A) 
is taken as the zero-point. The vertical line at $r=+13\arcsec$ corresponds to the kinematic center of 
the galaxy.
\label{fig:rot7292}}
\end{figure}

\begin{figure}
\vspace{3mm}
\centerline{\includegraphics[width=8.5cm]{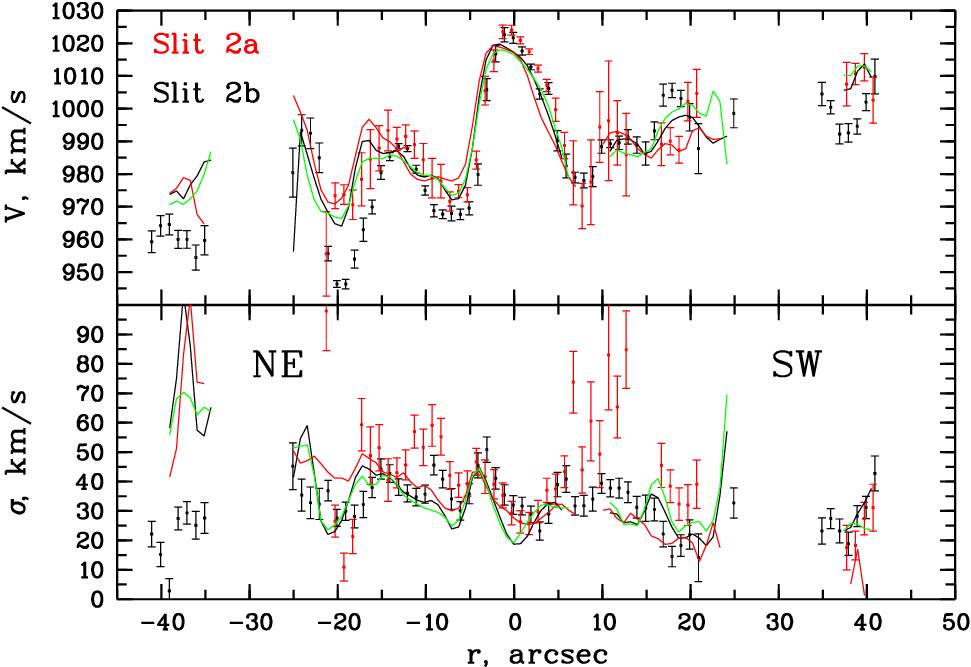}}
\caption{Radial velocity (top) and velocity dispersion (bottom) of HII along the second slit position. 
The black dots show the position of 2b (PA $= 28.48\degr$), the red dots -- the position of 2a 
(PA $= -151.52\degr$). Other designations are the same as in Fig.~\ref{fig:rot7292}.
\label{fig:rot7292b}}
\end{figure}

The measurements of the radial velocities of the gas in the galaxy obtained using the long-slit 
spectroscopy and scanning FPI are consistent with each other within errors (Figs.~\ref{fig:rot7292} 
and \ref{fig:rot7292b}). Both datasets show similar variations of the velocity dispersion along the 
slit, but the $\sigma^2$ measurements with the TDS are systematically higher than with the FPI, 
which is most likely due to two effects. First, the spectral resolution of the scanning FPI is one 
and a half times better, which is significant, since the observed line broadening due to the velocity 
dispersion is comparable to the width of the instrumental profile of the TDS 
($\sigma_{\rm lab} =50\pm1\,$km\,s$^{-1}$). Second, in the TDS observations, the radial velocity 
distribution inside the slit is averaged, which also somewhat broadens the line profiles. So, in the 
top panels of Figs.~\ref{fig:rot7292} and \ref{fig:rot7292b}, it is noticeable that the velocities 
determined from the velocity field constructed with the FPI by $\Delta V$ differ up to 5~km\,s$^{-1}$ 
within $1\arcsec$ (the black, red, and green curves).

The eastern end of the bar (the brightest HII region) with the velocity $V\approx1020$~\kms \, 
stands out kinematically. The velocity along the rest of the bar remains constant ($V=980-985$~\kms). 
The velocity dispersion of the gas $\sigma$ in the bar including its southeastern end is low, 
$20-40$~\kms. It corresponds to the $\sigma$ lower limit in NGC~7292 observed in some parts of 
the galaxy's disk. In the HII regions, $\sigma$ does not exceed 
30~\kms (Figs.~\ref{fig:ifpmaps}-\ref{fig:rot7292b}). This value is slightly larger than that 
obtained by \cite{law2022}, $\sigma=19\pm4$~\kms, for galaxies with the same star-formation rate 
(according to \cite{james2004}, in NGC~7292 SFR $=0.22\pm0.04 M_{\odot}$\,yr$^{-1}$), however, 
is consistent with the results by \cite{moiseev2015} for galaxies of comparable luminosity in the 
visible range and H$\alpha$.

\subsection{Parameters of the Disk Rotation from the Kinematics Data}

To measure the rotation curve of a thin flat disk, it is necessary to find the following parameters 
that determine its orientation and kinematics: the position angle of the line of nodes (of the major 
axis) PA$_0$, the inclination of the disk to the line of sight $i_0$, the coordinates of the 
rotation center, and $V_{SYS}$, $V_{ROT}$ -- the systemic velocity and the rotation velocity. At 
the same time, with each specific radius, the kinematic position angle of the major axis \pak \, and 
the inclination of the orbits of gaseous clouds to the perspective plane $i_{kin}$ may, for one 
reason or another, differ from PA$_0$ and $i_0$.

\begin{figure}
%\vspace{3mm}
\centerline{\includegraphics[width=8cm]{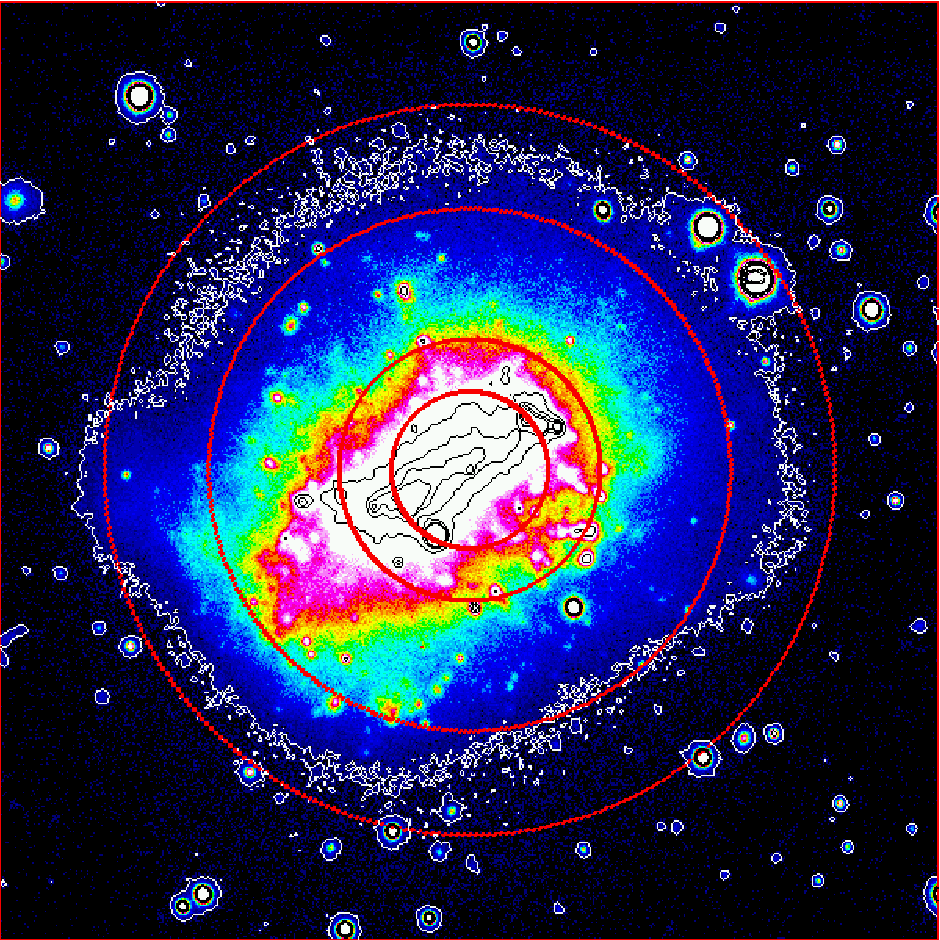}}
\caption{Deep image of NGC~7292 in the $r$ band of the Legacy Surveys before subtraction of the 
field stars. The radii of the circles are equal to $15\arcsec$, $25\arcsec$, $50\arcsec$, and 
$70\arcsec$. The external isophote given in white colour in $r$ corresponds to the level of 
$3\sigma$ higher than the sky background.
\label{fig:n7292r}}
\end{figure}

While \pak, and hence PA$_0$, is measured with confidence, as it characterizes the direction of 
the global radial velocity gradient, measuring the inclination angle $i_0$ is a problem. Analyzing 
the velocity field, one can conclude that this cannot be done, since the galaxy is in a position 
close to face-on ($i_0<40\degr$). The projected rotation velocity on the line of sight is not 
very large, while non-circular motions associated with the star-forming regions and gas flow in the 
bar make a very significant contribution to the observed situation. It is also difficult to 
establish the disk orientation from morphology, due to the fact that there is a strongly elongated 
bar in the inner part of the galaxy, and the outer regions are strongly perturbed 
(Figs.~\ref{fig:map1} and \ref{fig:n7292r}). Therefore, the position of the major axis along 
\pak$\approx250\degr$ seen from the radial velocity fields of HI (see Section~\ref{sect:intro} 
and HII does not correspond to the major axis of optical isophotes (see Section~\ref{sect:pp}) or 
distribution isodenses of HI. It is incorrect to determine the $i_0$ value from their ellipticity.

In order to estimate the disk inclination, we used the Tully-Fisher relation associating $V_{ROT}$ 
with the luminosity of the galaxy. We used the tilted-ring method to determine the maximum rotation 
velocity by analyzing the HI velocity field provided by \cite{biswas2022}. The angular resolution 
of the HI data is $30\arcsec\times23\arcsec$, and the spectral one corresponds to 6.6~\kms. Let us 
note that the deviations from the circular rotation in the southwestern protrusion of the HI disk 
do not exceed 5~\kms \, which is even smaller than those in some internal regions of the HI disk. 
The neutral gas disk is longer and the maximum in the rotation curve can be found at $r=60\arcsec$, 
where the ionized gas is no more observed. If we accept $M(B)=-16.7\pm1.0$ according to LEDA, from 
the calibration relations given in \cite{Rhee2005JASS...22...89R} for their RC sample, we obtain 
the supposed rotation curve amplitude 63~\kms. Fitting the radial velocity field using the 
tilted-ring model with different $i_0$ values showed that the required rotation velocities 
are achieved at $i_0=29\pm4\degr$.

The galactic disk rotation parameters were determined with the tilted-ring method adapted for 
analyzing the radial velocity field of the ionized gas \citep[the detailed description and main 
ratios are given in the papers by][]{Moiseev2014,Moiseev2021AstBu..76..316M}. According to the 
accepted parameters of the outer disk orientation PA$_0$, $i_0$, the velocity field was divided 
into narrow elliptical rings, the radial velocity distribution in each ring of the radius $r$ was 
fitted with a circular rotation model with the following parameters: $V_{SYS}$, $V_{ROT}$, \pak\, 
and $i_{kin}$. The center of rotation, determined from considerations of field radial velocity 
symmetry, coincides within error with a small HII region at the center of the bar 
\citep[region 5 from][located by $3.5\arcsec$ southeast of region C]{gusev2021}, therefore, in 
further analysis, it was fixed at this point with the coordinates 
$\alpha({\rm J}2000.0)=22^h28^m25.34^s$, $\delta({\rm J}2000.0)=+30\degr 17\arcmin 35.3\arcsec$. 
The measurement accuracy of the kinematic center position is approximately $\pm1\arcsec$. Moving 
the accepted position of the kinematic center to a greater distance (for example, to region C 
described in the Introduction and located at $3.5\arcsec$ from the chosen center) leads to 
noticeable (more than 10-15~\kms) variations in the systemic velocity along the radius, which are 
absent in the case of the rotation center we have adopted. Later, to obtain a more stable rotation 
model, the systemic velocity was fixed at the average value $V_{SYS}=988\pm1$~\kms.

\begin{figure}
%\vspace{2mm}
\centerline{\includegraphics[width=8cm]{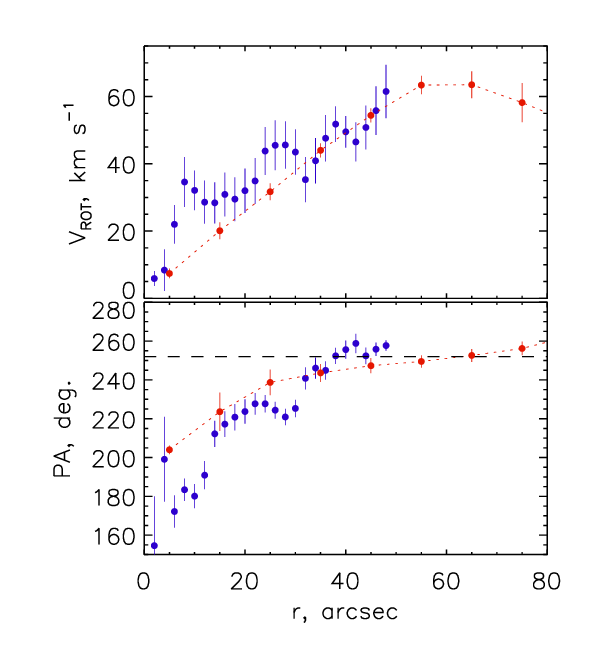}}
\caption{Parameters of the kinematic model: the ionized gas rotation curve (top) and the radial 
variations PA$_{kin}$ (bottom). The blue symbols show the ionized gas, the red mark the neutral 
hydrogen. The black dashed line shows the accepted orientation of the line of nodes of the gaseous 
disk.
\label{fig:rc}}
\end{figure}

The radial variations in the gas rotation parameters in the galaxy are shown in Fig.~\ref{fig:rc}. 
The orientation of the line of nodes defined as the average value of \pak{} at $r=35-45\arcsec$ 
is PA$_0=253\pm2\degr$. The rotation of the kinematic axis in the more internal regions of the 
galaxy is obviously associated with radial gas flows under the influence of the triaxial potential 
of the bar. The direct evidence here is the fact that the deviations of \pak{} from PA$_0$ occur 
in the opposite direction relative to the turn of the inner isophotes, whose major axis in the bar 
region is located along PA $\approx305\degr$ ($125\degr$); while in the outer regions, the 
isophotes gradually turn closer to the accepted PA$_0$ value (Figs.~\ref{fig:rc} and 
\ref{fig:fig_pae}). A detailed discussion of the technique for comparing photometric and kinematic 
PA in barred galaxies and references to the original papers are given in the review by 
\cite{Moiseev2021AstBu..76..316M}; a comparison with the results of numerical simulations is 
presented, for example, in the paper by \cite{Moiseev2000AstL...26..565M}.

Figure~\ref{fig:rc} also shows the results of applying the same rotation model to the velocity 
field in HI from \citet{biswas2022}. One can see a fairly good agreement between the kinematics 
of the ionized and neutral hydrogen taking into account the fact that the angular resolution in 
the 21-cm line is an order of magnitude lower and amounts to $30\arcsec\times23\arcsec$. Thus, 
measurements of \pak{} and $V_{ROT}$ in the H$\alpha$ line within $r<30\arcsec$ are sharper. 
The slow increase of \pak{} in HI for $r>40\arcsec$ is possibly indicative of the bending of 
the outer gaseous disk, in which the ionized gas is almost no longer observed.

\subsection{Photometric Parameters of the Disk}
\label{sect:pp}

\begin{figure}
\vspace{4mm}
\centerline{\includegraphics[width=8cm]{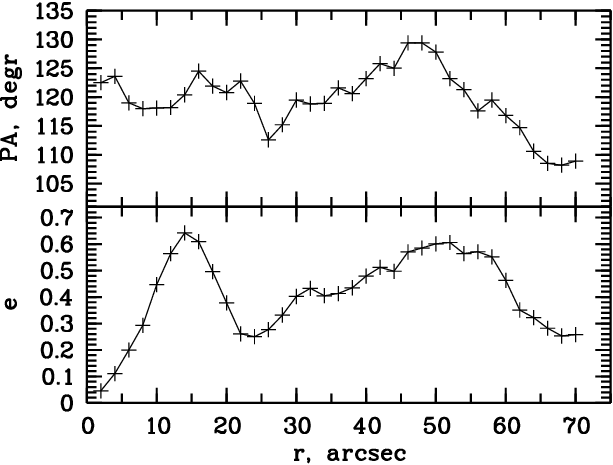}}
\caption{Radial proﬁles of the position angle PA and ellipticity $e$ obtained for the image of 
NGC~7292 in the $r$ band.
\label{fig:fig_pae}}
\end{figure}

For comparison, using the data on the coordinates of the kinematic center of NGC~7292 and the 
image of the galaxy in the $r$ band from the Legacy Surveys (Fig.~\ref{fig:n7292r}), we present 
the obtained radial distribution of the photometric position angles PA and the isophote 
ellipticity $e\equiv1-b/a$ (Fig.~\ref{fig:fig_pae}). The parameters PA and $e$ were calculated 
using the SURFPHOT software package in the MIDAS environment. The field stars were previously 
removed from the image shown in Fig.~\ref{fig:n7292r}. The star-forming regions (the HII ones) 
of the galaxy were not subtracted from the image. As can be clearly seen from 
Figs.~\ref{fig:n7292r} and \ref{fig:fig_pae}, they make a significant contribution at 
galactocentric distances $r<30-35\arcsec$ without affecting the determination of the parameters 
for the outer-disk regions of NGC~7292.

Figure~\ref{fig:fig_pae} shows a complex structure of NGC~7292. In the bar region ($r<15\arcsec$), 
the ellipticity increase from 0.05 to 0.65 is observed at the constant position angle 
(PA $=118-123\degr$). In the range of $r=15\arcsec-25\arcsec$, the ellipticity drops sharply 
to 0.3 and the position angle decreases to $113\degr$. This region corresponds to the brightest 
area of the disk actively emitting in the H$\alpha$ line and limited by the chains of HII regions 
(see galaxy images in Figs.~\ref{fig:map1}, \ref{fig:ifpmaps}, and \ref{fig:n7292r}). In the 
range of the galactocentric distances of $25\arcsec-50\arcsec$, a disk of moderate brightness 
can be observed with its symmetry center displaced relative to the kinematic center of NGC~7292 
and located near the southeast end of the bar of the galaxy (Fig.~\ref{fig:n7292r}). In this 
region, the isophote position angle reaches the maximum at $129.5\degr$, and the ellipticity 
increases to 0.6. Both parameters, PA and $e$, at $r=50\arcsec$ are close to the values obtained 
for $r=15\arcsec$. Finally, in the outermost part of the galaxy, at $r=70\arcsec$ (2.3~kpc), 
the ellipticity and position angle again decrease to the values close to those obtained at 
$r=25\arcsec$ (Fig.~\ref{fig:fig_pae}).

As can be clearly seen, the parameters PA and $e$ determined from the photometric data do not 
agree with those obtained from the velocity field analysis.

\section{DISCUSSION}

Figure~\ref{fig:ifpmaps} shows the velocity field in the circular rotation model (more 
precisely, quasi-circular, since \pak \, varies with $r$) and residual velocities 
$V_{res}$ after subtracting the model from observations. There are several features 
in their distribution. First of all, these are high moduli of $V_{res}$ in two giant star-forming 
regions at the ends of the bar. This symmetric arrangement may be due to the fact that the 
residual velocities are associated with non-circular (usually, radial) movements in the bar, which 
are not fully accounted for by our quasi-circular rotation model. But it is strange that the 
amplitude of velocities in the regions symmetrical with respect to the center differs by a factor 
of two: $V_{res}$ is --10...--15~\kms \, in the northwestern but reaches +30...+37~\kms \, in 
the southeastern region (region~A). Then the velocity of the radial motions will be, in terms of 
the disk plane, $V_{res}/\sin i=$ 23--35~\kms \, and 71--87~\kms, respectively. And if the first 
estimate looks reasonable, in the second case the velocity of the gas flow to the center exceeds 
the hyperbolic one at the given radius ($V_{ROT}\approx30$~\kms \, at $r=20\arcsec$, 
Fig.~\ref{fig:rc}) which is doubtful.

It is possible that such high peculiar velocities in the southeastern region designated as A 
(see Fig.~\ref{fig:ifpmaps}) are not related to the radial direction but to some other, for 
example, to the vertical motions of the gas in the disk. The gas outflows from the star-forming 
regions associated with the gaseous envelope expanding under the action of supernova outbursts and 
stellar winds could be the most obvious reason. But in this case, it is not clear why we observe 
only the moving-away part of the envelope ($V_{res}>0$) and not the one expanding towards the 
observer. Moreover, the velocity dispersion map in the H$\alpha$ line (Fig.~\ref{fig:ifpmaps}), 
which is consistent with the long-slit spectroscopy data (Fig.~\ref{fig:rot7292}), shows a pattern 
typical of dwarf galaxies with the ongoing star formation \citep{MoiseevLozinskaya2012}, namely: 
$\sigma$ has minimum values at the centers of the bright HII regions, since here the main emission 
comes from denser gaseous clouds. At the same time, $\sigma$ is higher on their periphery, since 
here we can see the emission of the diffuse gas which is more sensitive to the star-formation 
perturbation. If large values of $V_{res}$ in the southeastern region were associated, for example, 
with a breakthrough of the expanding gas envelope, then it would be expected to observe a peak in 
the velocity dispersion. But this is not observed.

A typical situation associated with the interstellar medium perturbation caused by the star 
formation is observed along the major axis of the galaxy in the southwestern and northeastern parts 
of the disk, where regions of high $\sigma$ have a low brightness in the H$\alpha$ line and 
relatively high values of $V_{res}$. At the same time, at least some of the non-circular gas 
motions near the large star-forming complexes is most likely associated with the tidal perturbation 
of the disk after the last interaction. Is it possible that the area southeast of the center is part 
of a companion that merged with NGC~7292? According to \cite{gusev2021}, this complex has the 
lowest internal extinction of all the studied HII regions in NGC~7292 determined from the Balmer 
decrement. This may testify in favor of the assumption that it enters the gaseous disk of the galaxy 
from the side of an observer from the Earth. On the other hand, the star-formation complex does not 
stand out by the oxygen abundance (its O/H $=8.26\pm0.05$~dex corresponds to the average of the 
galaxy), and the position of the complex in the characteristic diagrams does not indicate the 
presence of the HII shock excitation mechanisms \citep{gusev2021}. Thus, the currently available 
observed data do not allow one to unambiguously identify the structures associated with a companion 
that has been swallowed up by the main galaxy.

\begin{figure}
%\vspace{2mm}
\centerline{\includegraphics[width=8cm]{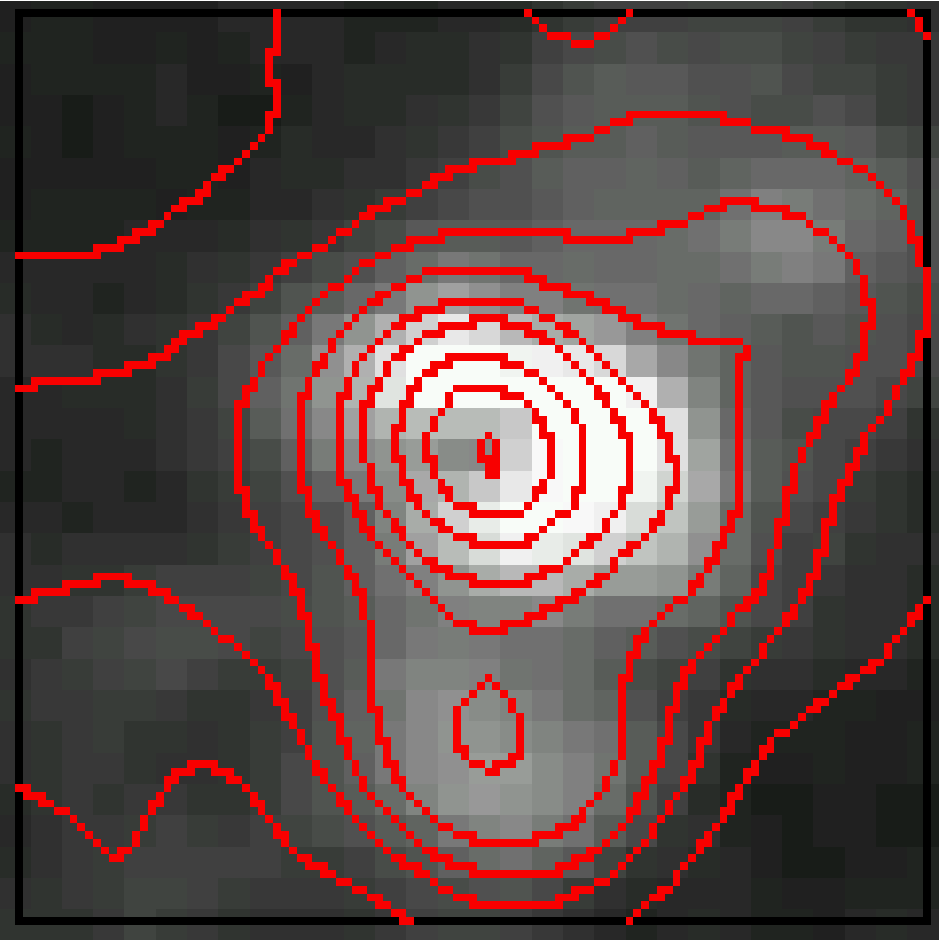}}
\caption{Image of the brightest HII region of NGC~7292 (region A at the southeastern end of the 
bar) in the H$\alpha$+[NII] line with the superimposed isophotes in the $U$ band. The image size 
is $8\arcsec\times8\arcsec$ (260~pc). North is on the top, East is on the left. We used the 
unpublished $U$ data obtained with the 1.5-m telescope of the Maydanak Observatory (Uzbekistan), 
and in the H$\alpha$ line with the continuum subtracted obtained with the 2.5-m telescope of the 
CMO SAI MSU from the personal archive of one of the authors. The observations in the continuum 
(the filter with $\lambda_{\rm eff}=6427$\AA, $\Delta\lambda=122$\AA) were carried out together 
with the observations in the H$\alpha$+[NII] line.
\label{fig:sfr7292}}
\end{figure}

This complex is at an evolutionally advanced stage of the star formation. Its N/O ratio is the 
highest among all the studied HII regions \citep{gusev2021} and indicates the enrichment of the 
surrounding interstellar medium with nitrogen of secondary origin. The morphology of the complex 
also evidences the decrease (termination within the model of a single starburst) of the 
star-formation rate in the modern age. Figure~\ref{fig:sfr7292} clearly shows that the 
H$\alpha$-emission region forms a ring of a diameter of 50-80~pc surrounding the star complex. 
According to \cite{whitmore2011}, this stage of evolution (''young cluster'') occurs approximately 
6-8~Myr after the starburst, when the first supernovae sweep out the gas to the periphery of the 
star complex. Note the asymmetry of the HII envelope: the maximum H$\alpha$ emission is observed 
to the west (from the side of the bar) of the complex, where the gas density is apparently the 
highest.

If we assume that the star formation in this region is stimulated by an infall of a companion 
(gaseous or stellar-gaseous), then it should have occurred relatively recently, on time scales 
of about 10~Myr ago. This is indicated by: the age of the stellar population of complex A; the 
spatial scale of perturbations of the gaseous disk of NGC~7292 (the region of abrupt changes in 
radial velocities from 1020 to 980~\kms\, is $5-8\arcsec$; 
Figs.~\ref{fig:ifpmaps}-\ref{fig:rot7292b}) which corresponds to the same time scale of 
7-8~Myr with the observed $\sigma\approx30$~\kms\, in this region); the distance traveled by 
a possible companion with the projected radial velocity relative to the galaxy equal to 
$40$~\kms, for a time of 6-8~Myr, in order of magnitude, the effective thickness of the NGC~7292 
disk is about 200~pc. From the available data, we cannot say whether this local perturbation 
(a possible infall of a companion) is associated to longer-lived in time and larger-scale in 
distance evidence of tidal perturbations such as the bending of the HI outer disk beyond the 
optical radius, the asymmetry in the HI distribution, and the distorted shape of the external 
optical isophotes, or it is independent.

The galaxy has not previously been studied kinematically in detail. The newly observed data we 
have obtained show the peculiar, very difficult to interpret features of gas motion in the disk 
of NGC~7292. Having considered all possible cases above, we believe that the most realistic way 
to explain the radial velocity anomaly at the southeastern end of the bar (the brightest HII region 
in the galaxy previously taken as its center) is the infall of an outer companion.

\section{CONCLUSION}

The analysis of the velocity fields of the ionized and neutral hydrogen showed that the kinematic 
center of NGC~7292 is located at the center of the bar, northwest of the photometric center of 
the galaxy (the southeastern end of the bar) which has been previously taken as the center of 
NGC~7292. In addition to the circular rotation, the radial motions associated with the bar play 
a significant role in the kinematics of the gaseous disk. Compared to these non-circular motions, 
the observed perturbation of the gaseous disk by the star formation is not so strong: the 
differences from the circular rotation in the HII regions do not exceed those caused by the radial 
flows in the bar, and the velocity dispersion increases only in the diffuse ionized gas closer to 
the disk periphery. The velocity dispersion distribution does not contain any envelope structures 
characteristic of a number of dwarf galaxies with a starburst 
\citep{MoiseevLozinskaya2012,Gerasimov2022MNRAS.517.4968G}. It is possible that some of the 
non-circular motions (primarily at the southeastern end of the bar -- the brightest HII region) 
may be associated with the consequences of a merger with a dwarf companion or the capture of 
an external gaseous cloud. However, so far we have not managed to find direct evidence of this 
interaction. The presence of tidal perturbations, such as the bending of the outer HI disk 
beyond the optical radius, the asymmetry in the neutral hydrogen distribution, and the distorted 
shape of the outer optical isophotes, can be related to a possible infall of a companion at 
the southeastern end of the bar, or independent of it.

\section*{ACKNOWLEDGMENTS}

The authors are grateful to the reviewer for valuable and useful comments. We thank 
E.~I.~Shablovinskaya and E.~A.~Malygin, who carried out observations with the 6-m SAO RAS 
telescope, A.~V.~Dodin, who did the primary reduction of the long-slit spectroscopy data, and 
Prerana Biswas, who kindly provided the HI maps obtained with the GWRT telescope. The open data of 
the HyperLEDA database (http://leda.univ-lyon1.fr) and NASA/IPAC Extragalactic Database 
(http://ned.ipac.caltech.edu) were used in the paper. We used the public data from the Legacy Survey 
(http://legacysurvey.org) consisting of three separate and complementary projects: the Dark Energy 
Camera Legacy Survey (DECaLS; Proposal ID No~2014B-0404; PIs: David Schlegel and Arjun Dey), the 
Beijing-Arizona Sky Survey (BASS; NOAO Prop. ID No~2015A-0801; PIs: Zhou Xu and Xiaohui Fan), and 
the Mayall $z$-band Legacy Survey (MzLS; Prop. ID No~2016A-0453; PI: Arjun Dey). DECaLS, BASS, and 
MzLS together include the data obtained, respectively, with the Blanco telescope, Cerro Tololo 
Inter-American Observatory, NSF's NOIRLab; the Bok telescope, Steward Observatory, University of 
Arizona; and the Mayall telescope, Kitt Peak National Observatory, NOIRLab. The Project Legacy 
Survey was honored to receive permission to perform the astronomical research on Iolkam Du\'{a}g 
(Kitt Peak), a mountain of special significance to the Tohono O\'{o}dham people.

\section*{FUNDING}

The spectroscopic observations at the Caucasian Mountain Observatory of SAI MSU and the analysis 
of the long-slit spectroscopy data were supported by the RFBR grant No. 20-12-00080 and by the 
Interdisciplinary Scientific and Educational School of the Lomonosov Moscow State University 
''Fundamental and Applied Space Research''. The observed data were partially obtained with the 
2.5-m telescope of the Caucasian Mountain Observatory of SAI MSU. The development of the 
instrumental base of the observatory is carried out with the support of the Development Program 
of the Lomonosov Moscow State University. Observations with the 6-m SAO RAS telescope are supported 
by the Ministry of Science and Higher Education of the Russian Federation. The renovation of the 
instrumental base is carried out within the framework of the ''Science and Universities'' national 
project. The analysis of the ionized gas kinematics was carried out within the framework of the 
state assignment of SAO RAS approved by the Ministry of Science and Higher Education of the 
Russian Federation.

\section*{CONFLICT OF INTEREST}

The authors declare no conflict of interest regarding the publication of this paper.

\vspace{4mm}

{\it The paper was translated by N.~Oborina}

\end{document}